\def\Draft{1}% draft=1 or no draft = 0
\def\Draft{0}% draft=1 or no draft = 0
\newcommand{\AuthorLuP}{Laurent U.~Perrinet}%
\newcommand{\AuthorWT}{Wahiba Taouali} %
\newcommand{\AuthorGB}{Giacomo Benvenuti}%
\newcommand{\AuthorPW}{Pascal Wallisch} %
\newcommand{\AuthorFC}{ Fr\'ed\'eric Chavane} %
\newcommand{\AddressA}{Institut de Neurosciences de la Timone, CNRS, Aix-Marseille Universit\'e, Marseille, France}% - Marseille, France
\newcommand{\AddressB}{Center for Neural Science, New York University, New York, USA}%
\newcommand{\LongAddressA}{
Institut de Neurosciences de la Timone (UMR 7289),
CNRS, Aix-Marseille Universit\'e \\

27, Bd Jean Moulin,
13385 Marseille Cedex 05,
France}
\newcommand{\PhoneWT}{+33.662 866 453}%
\newcommand{\Website}{http://invibe.net/LaurentPerrinet}%
\newcommand{\EmailWT}{wahiba.taouali@univ-amu.fr}%
\newcommand{\Title}{
Testing the Odds of Inherent versus Observed Over-dispersion in Neural Spike Counts
}%
\newcommand{\RunningTitle}{Odds of Inherent versus Observed Over-dispersion}
\newcommand{\Abstract}{
The repeated presentation of an identical visual stimulus in the receptive field of a neuron may evoke different spiking patterns at each trial. Probabilistic methods are essential to understand \xchanged{its}{the} functional role \xadded{of this variance} within the neural activity. In that case, a Poisson process is the most common model of trial-to-trial variability. \xchanged{However}{For a Poisson process}, the variance of the spike count is constrained to be equal to the mean, irrespective of \xchanged{measurement's duratio}{the duration of measurements}. Numerous studies have shown that this relationship does not generally hold. Specifically, a majority of electrophysiological recordings show an ``{\em \xchanged{overdispersion}{over-dispersion}}'' effect: Responses that exhibit more inter-trial variability than expected from a Poisson process alone. A model that is particularly well suited to quantify \xchanged{overdispersion}{over-dispersion} is the Negative-Binomial distribution model. This model is \xchanged{largely applied and studied}{well-studied and widely used} but has only recently been applied to neuroscience. In this paper, we address three main issues. First, we describe how the Negative-Binomial distribution provides a model apt to account for overdispersed spike counts. Second, we quantify the significance of this model for any neurophysiological data by proposing a statistical test, which quantifies the odds that \xchanged{overdispersion}{over-dispersion} could be due to the limited number of repetitions (trials). We apply this test to three neurophysiological tests along the visual pathway. Finally, we compare the performance of this model to the Poisson model on a population decoding task. \xchanged{This shows that more knowledge about the form of dispersion tuning is necessary to have a significant gain, uncovering a possible feature of the neural spiking code.}{We show that the decoding accuracy is improved when accounting for over-dispersion, especially under the hypothesis of tuned over-dispersion.}
}%
\newcommand{\Keywords}{Spike counts, \xchanged{overdispersion}{over-dispersion}, Negative-Binomial distribution, Decoding, Tuning function.}%
\newcommand{\Acknowledgments}{%
WT, FC and LP were supported by EC FP7-269921, ``BrainScaleS'', GB by FACETS ITN project (EU funding, grant number 237955), a 'Marie-Curie Initial Training Network' and  PW was supported by F32-EY019833 from the NIH. We wish to thank Philippe Foundation for a travel grant support to F.C. The authors are indebted to T. Movshon and his laboratory for giving the opportunity to  F.C. to perform MT experiments at CNS/NYU. } %Code and supplementary material :\\ \url{\Website/Publications/Taouali15} %
\documentclass[a4paper, 10pt, twoside]{article}
\usepackage[english]{babel}%
\usepackage[autostyle]{csquotes}
\usepackage{hyperref}
\if 1\Draft
\usepackage{setspace}
\usepackage{lineno}
\fi
\usepackage{url}
\usepackage{graphicx}
\usepackage{geometry}
\usepackage{tikz}
\usetikzlibrary{positioning}
\usepackage[natbib=true,
			style=nature, %numeric-comp,
      		sortcites=true,
			abbreviate=true,
			maxcitenames=2,
			maxnames = 5,
			doi=true,
			url=true,
			isbn=false,
			eprint=false,
			texencoding=utf8,
			bibencoding=latin1,
			backend=bibtex,
			]{biblatex}%
\AtEveryBibitem{
  \clearfield{month}
  \clearfield{day}
  \clearfield{note}
  \clearfield{comment}
}
\usepackage{fancyhdr}
\title{\Title}
\date{1 January 2016}

\author{{\AuthorWT}$^{1}$, \AuthorGB$^{1}$, \AuthorPW$^2$,\\
 \AuthorFC$^1$ and \AuthorLuP$^1$\\
}
\newcommand{\xchanged}[2]{#2}
\newcommand{\xremoved}[1]{}
\newcommand{\xadded}[1]{#1}

\begin{document}
\if\Draft1
\linenumbers
\fi
\maketitle

\paragraph{Institutions}~~\\

$^{1}$ \AddressA\\

$^{2}$ \AddressB\\

\paragraph{Abbreviated title for the running head}~~\\

\RunningTitle

\paragraph{Corresponding author contact}~~\\

Name : \AuthorWT \\

E-mail: \EmailWT\\

Phone number : \PhoneWT\\

Physical address: \LongAddressA \\

%----------------------------------------------------------------------------------------------%
\begin{abstract}
\Abstract
\end{abstract}
%\smallskip
%\indent\indent\indent 
\paragraph{BibTex entry}~~\\
\begin{verbatim}
@article{Taouali15,
    title = {Testing the Odds of Inherent versus Observed 
             Over-dispersion in Neural Spike Counts},
    author = {Taouali, Wahiba and
              Benvenuti, Giacomo and
              Wallisch, Pascal and
              Chavane, Frederic and
              Perrinet, Laurent U.},
    journal = {Journal of Neurophysiology},
    year = {2015},
    doi = {10.1152/jn.00194.2015},
    issn = {1522-1598},
    volume = {115},
    number = {1},
    pages = {434-444},
    url = {http://dx.doi.org/10.1152/jn.00194.2015},
    url = {http://invibe.net/LaurentPerrinet/Publications/Taouali15},
    pmid = {26445864},
    publisher = {American Physiological Society},
}
\end{verbatim}
\if\Draft1
\newpage 
\doublespacing
\fi
%--------------------------------------------------------------------------------------------------%
\section*{Introduction}
\begin{figure}[ht]
\includegraphics[width=\textwidth]{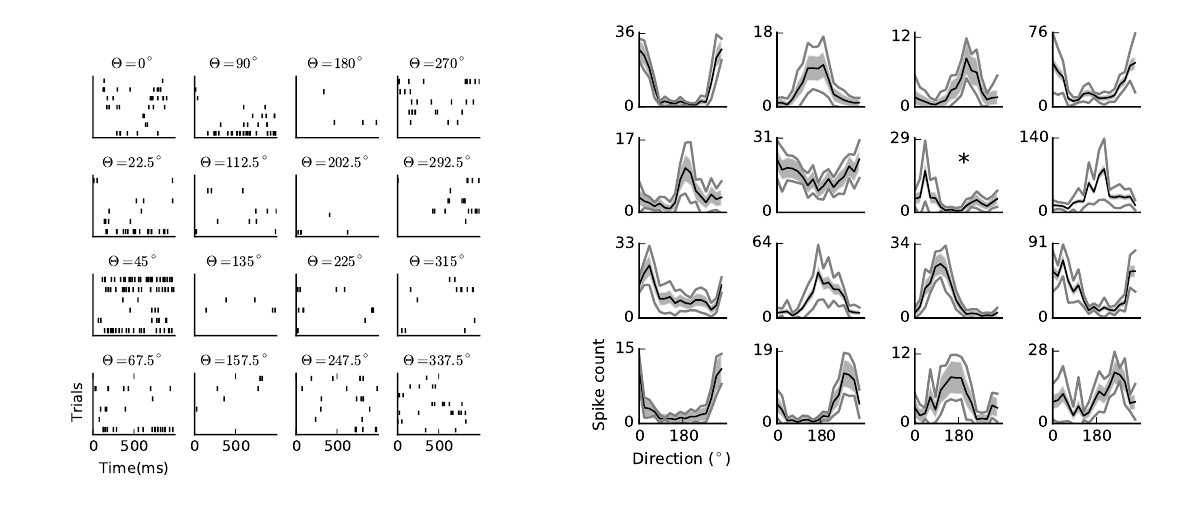}
\caption{{\bf Trial-to-trial variability of neural responses.}
(Left)~We show the spiking activity of a representative cell of the area MT of a macaque monkey (see Methods section) in response to oriented drifting gratings. Each subplot corresponds to a raster plot of the responses to 16 equally distributed motion directions. There is a high trial-to-trial variability in the spiking responses. %
(Right)~To quantify this variability, we plot \xchanged{for a population of 16 cells from the same area the directional tuning curves in response to the same 16 directions (mean spike count, black lines) along with the range of {$\pm$} the standard deviation (red lines)}{the directional tuning curves in response to the same 16 directions (mean spike count, black lines) along with the range of {$\pm$} the standard deviation (red lines) for a population of 16 cells from the same area}. The subplot with a star corresponds to the representative cell on the left. The area corresponding to the mean {$\pm$} the standard deviation expected under Poisson distribution hypothesis {($\sigma=\sqrt \mu$)} is plotted in blue. In this experiment, most of the cells show a higher variability than expected from the Poisson model.}
\label{fig:obs}
\end{figure}

%--------------------------------------------------------------------------------------------------%
One of the most common ways to study the central nervous system is to use a stimulus-response paradigm. A system, which could be a neuron or a population of neurons, is provided with an input (a stimulus) and the resulting state change (response) of the system is quantified in order to better characterize functional associations. A main problem with this approach is that the system's response to repeated presentations of an identical visual stimulus often exhibits high variability, particularly in the spike count statistics (see Figure~\ref{fig:obs}). As a consequence, statistical methods are essential to characterize the information carried by neural populations. 
%The application of Fano's method to neuroscience (and indeed biology) happened very late. For cases of variability, I like to  (also) cite the original papers by Fano, where he struggled with similar issues in physics. To my knowledge, this is the first time anyone did: http://journals.aps.org/pr/abstract/10.1103/PhysRev.72.26

The Fano factor~\citep{Fano47} is commonly used to quantify the variability of spike counts~\citep{Kara00, Alitto11, Zhuang13,Tolhurst83, Softky93, Shadlen98, Koch99}. It is defined by $FF(\Delta t) = \frac{\sigma^2(\Delta t)}{\mu(\Delta t)}$ where $\mu$ and $\sigma^2$ are the average and the variance over multiple trials of the number of spikes in some time window of length $\Delta t$, respectively. %: Poisson model
This variability is most often characterized by the Poisson spiking Model (PM). By the definition of this model, the spike count follows a homogeneous Poisson process with a rate parameter ${\lambda}$ (the expected number of spikes that occur per unit of time). If $k$ is the observed number of spikes fired by a neuron in response to a stimulus %defined by a variable $\theta$%
in a time window $\Delta t$,  the probability of such an observation is given by a Poisson distribution of parameter $f=\lambda \Delta t$, which gives the expected number of spikes per sample in the considered time window:
\begin{equation}
P(k) = \frac{f ^{k}e^{-f}}{k!}
\end{equation} 
It follows that $\mu=f$ and $\sigma^2=f$ and thus that the Fano factor under the Poisson model has a value of $1$, regardless of $\lambda$ or $\Delta t$. 
%%%%
Such a property is in accordance with a wide range of observations of variability in cortical responses to visual stimuli~\citep{Shadlen98,Softky93,Britten93,Buracas98,McAdams99} showing variance-to-mean ratios of spike counts that equal or just exceed unity. However, for a large majority of reported evoked responses, the variability is higher than expected from a Poisson model, \xchanged{specially}especially in mammalian visual cortex \citep{Heggelund78, Dean81, Schiller76, Tolhurst83, Vogels89}, \xadded{whereas} Sub-Poisson variability has been observed only under specific conditions \citep{Kara00}.  

In fact, the variability of neural responses depends on different parameters. Some studies reported that the Fano factor values can vary depending on the brain region~\citep{Kara00, Gur06, Kayser10}. Indeed, neural responses in higher cortical areas~\citep{Maimon09} or in other non-cortical sensory areas such as the retina~\citep{Berry97} and the thalamus~\citep{Kara00} show substantially lower variability. It has been also shown that some neural features such as the refractory period~\citep{Kara00, Berry97} lead to a decrease of variability. Moreover, the nature of the stimulus itself could contribute to the variability of neuronal responses. It has been suggested that increasing the complexity of the stimuli by using natural-like statistics increases the reliability~\citep{Baudot13,Haider10,Borst99}. In addition, it has been reported that the variability of evoked responses is not stationary even during a trial time-course and decreases after the presentation onset of most sensory stimuli \citep{Churchland10}. Finally, a recent study~\citep{Nienborg12} on behaving animals \xchanged{focused on}{explored} the role of decision and other cognitive processes in shaping variability.

%--------------------------------------------------%
%: -observations
% ------------------------------------------------- %
%-------------------------------------------------%
%: fig:obs
%------------------------------------------------- %
%------------------------------------------------- %
In this paper, we mainly explore \xchanged{overdispersion}{over-dispersion}, that is, cases where there is greater variance in the spike count than might be expected by the Poisson model. Our hypothesis is that the Poisson process only accounts for the intrinsic noise induced by the spiking mechanism itself. Indeed, under the Poisson process hypothesis, the trial-to-trial variability only results from the stochastic properties of the neuron~\citep{Carandini04, Mainen95, Schneidman98}. Thus, the PM can not account for other significant sources of variability including hidden contextual variables such as attention, varying cortical states~\citep{Arieli96,Buracas98,Tsodyks99,Kenet03}, perceptual effects~\citep{Ress03} and overt behaviors (such as the precise eye position)~\citep{Gur97}. Thus, we consider an alternative compound distribution, called Negative-Binomial, that generalizes the Poisson distribution with a dispersion parameter that directly controls the ratio between the mean and \xremoved{variance }variance. Such a doubly-stochastic model was recently proposed to model the variability of neural dynamics showing more complex variability than what is expected from a Poisson model~\citep{Churchland11, Ponce13, Ponce10,Goris14}. Our aim is to further explore the potential suitability of this doubly-stochastic model to account for response variability in neurophysiological data.

%: Outline:
The paper is organized as follows. First, we describe the Negative-Binomial encoding model (NBM) and two possible implementations of \xchanged{overdispersion}{over-dispersion} sources. Second, we propose a statistical test for evaluating whether the observed \xchanged{overdispersion}{over-dispersion} is significantly better described by the Negative-Binomial model than the Poisson model. To assess the generality of this model, we then apply this test to three data sets of spiking responses in different animals, different states and different visual stimuli: in the lateral geniculate nucleus (LGN) neurons of anaesthetized mice presented with drifting gratings, in the primary visual cortex (V1) of awake macaque monkeys viewing oriented moving bars and in the middle temporal cortex (MT) of anaesthetized macaque monkeys presented with drifting gratings. Finally, we compared the efficiency of PM and NBM models on a population decoder using the most dispersed data set (MT).  

% ------------------------------------------------- %% ------------------------------------------------- %% 

%---------------------------------------------------------------------------------------------------------------- %
\section{ The Negative-Binomial model: encoding of inter-trial \xchanged{overdispersion}{over-dispersion}}
%--------------------------------------------------------------------------------------------------%% ------------------------------------------------- %
%We illustrate this approach with the negative binomial regression model, which is the best known and most widely available Poisson-based regression model that allows for \xchanged{overdispersion}{over-dispersion}.

Several models have been developed to account for over-dispersion such as generalized Poisson~\citep{Consul73}, zero-inflated Poisson~\citep{Lambert92, Miaou94, Shankar97} or quasi-Poisson~\citep{ Mccullagh89, Wedderburn74} models. However, \xadded{only a} few such as the Negative-Binomial model (NBM) are well known and widely available. \xremoved{Indeed, }The Negative-Binomial model has been \xremoved{largely} studied and tested in different fields such as epidemiology~\citep{Byers03}, accident statistics~\citep{Poch96}, biology~\citep{Anscombe50, Bliss53}, etc. However, it has only recently been applied to evoked neural responses~\xchanged{\protect \citep{Goris14}}{\protect \citep{Scott12,Goris14}} where the variability was traditionally supposed to be well modeled by a Poisson model (PM).    

A Negative-Binomial model depends on a distributional form equivalent to a compound stochastic process. \xadded{It has recently been represented as a Polya-Gamma mixture of normals \protect \citep{Polson12}}, but its standard representation is the Gamma-Poisson mixture. It corresponds to a doubly stochastic process where the Poisson process has a probability distribution $\mathcal{P}$ with a parameter $\lambda$ which is itself a random variable generated from a Gamma distribution $\mathcal{G}(\phi,\beta)$ with a shape parameter $\phi$ and a scale parameter $\beta$. When we consider a random variable $x$ generated by this compound process (i.e. $x \sim \mathcal{P}(\lambda)$ with $\lambda  \sim \mathcal{G}(\phi,\beta)$), the resulting distribution of this variable corresponds to a Negative-Binomial distribution (NB) which is usually defined by its shape parameter $\phi$ and its scale parameter $p=\frac 1{\beta+1}$:
\begin{equation}
 x \sim \mathcal{NB}(\phi, p=\frac 1{\beta+1})
 \label{eq:NB}
\end{equation}

Using the Negative-Binomial model, the probability of getting $k$ spikes from a given cell with parameters $\phi$ and  $p$ in response to a specific stimulus is given by:
 \begin{equation}
P(k; \phi, p)=\frac{\Gamma(k+\phi) \cdot p^{\phi}\cdot(1-p)^{k}}{\Gamma(\phi)k!}\end{equation}
We deduce the mean ($\mu$), the variance ($\sigma$) and the Fano factor (${FF}$) as:
\begin{equation}
\mu= \phi \cdot \beta \mathrm{, ~}  \sigma^2  = \mu + \frac{\mu^2}{\phi}\mathrm{~and~{FF}} =  \frac{\sigma^2}{\mu}=1 + \frac{\mu}{\phi}
\end{equation}
This distribution approaches the Poisson probability of parameter $\mu$ for large values of $\phi$, but allows to reach larger-than-Poisson variances for small $\phi$ values. We end up with a multivariate distribution  \xadded{that is} - in practice - \xremoved{is} driven by two parameters: the mean spike count ($\mu$) and the parameter ($\phi$) that we call the {\em ``inverse-dispersion parameter"} \footnote{ The inverse of the parameter $\phi$, usually denoted by $\alpha$, has been often called in the literature as the {\em ``dispersion parameter"}. It has been occasionally named the {\em ``\xchanged{overdispersion}{over-dispersion} parameter"} because it does not account for FF values smaller than one.}.

 \xadded{Under the NBM hypothesis, the distribution of inter-spike intervals (ISIs) is given by a Loamax distribution (equivalent to an exponential distribution with a random parameter generated from a Gamma distribution). The mean and the variance of such a distribution are given by the following expressions: mean(ISIs) = $\frac {\beta} {(\phi-1)} $ defined for $\phi >1$, variance(ISIs) = $\frac {\beta^2 \phi} {(\phi-1)^2(\phi-2)} $ defined for $\phi >2$. The resulting squared coefficient of variation is given by :
 \protect \begin{equation} 
   CV^2 = \frac {\phi} {(\phi-2)}=1+\frac {1} {\frac \phi 2-1}
 \protect \end{equation}
defined for $\phi >2$. $CV^2$ is higher than 1, which implies than a Negative-Binomial process is more irregular than a Poisson process and approaches Poisson regularity for large values of $\phi$. However, unlike under the Gamma distribution hypothesis made in \protect~\citep{Nawrot08}, the Negative-binomial model do not allow us to systematically relate over-dispersion to the positive deviation from the null-hypothesis $FF=CV^2$.} Thus, to better qualify this \xchanged{overdispersion}{over-dispersion} parameter, we need to focus on how to generate the Gamma input drive that feeds the Poisson generator.

%------------------------------------------------- %
%: fig:model1
\begin{figure}[t]
\includegraphics[width=\textwidth]{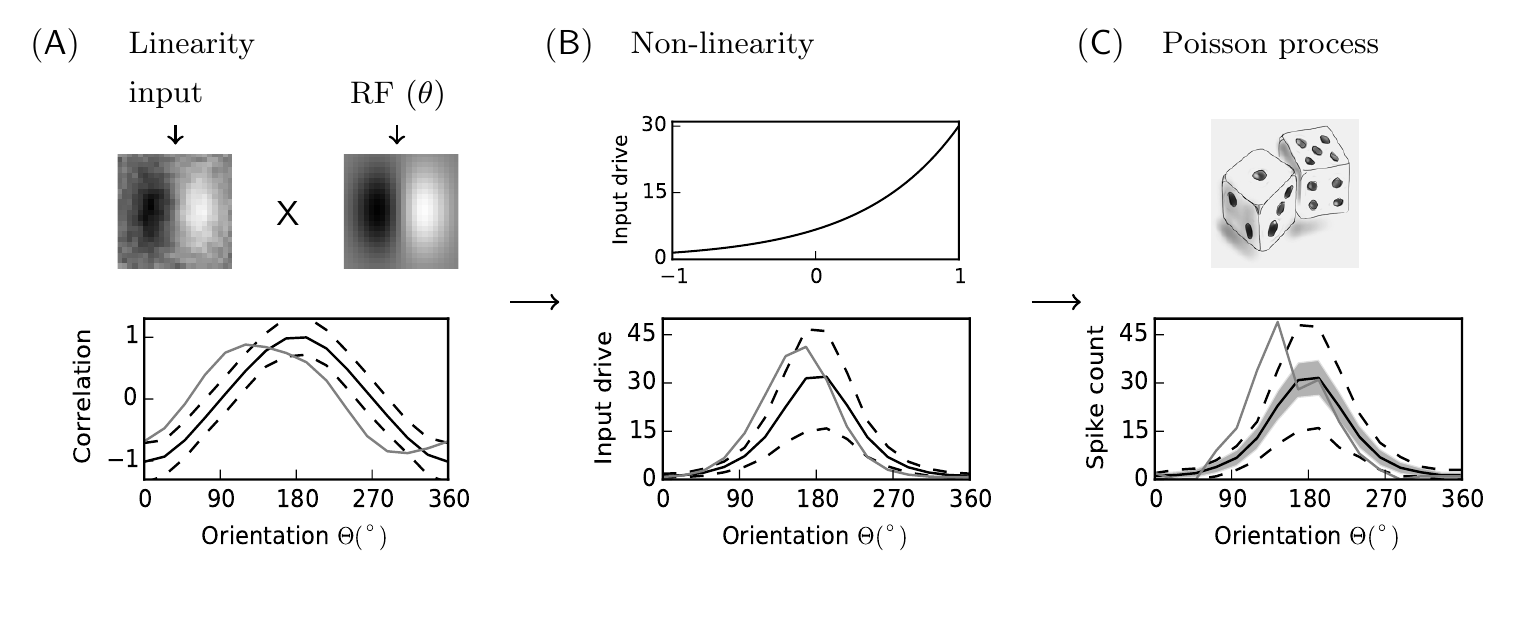}
\caption{{\bf Over-dispersion may result from extrinsic sensory noise\xadded  {: NBM as a mixture of Poisson and LogNormal distributions.}} We show here a simple linear / non-linear spiking model with three processing steps: (A)~The correlation between an noisy Gabor image (input) and a population of Gabor filters at different orientations models the linear processing in the receptive field (RF) of a visual neuron. The green, black and red lines correspond respectively to one trial (simulation), the average and the average $\pm$ the standard deviation (within 1000 trials). (B)~This input is transformed using a static, exponential non-linearity multiplied by a scaling factor corresponding to the maximum spike count. \xadded{The input drive resulting from a Gaussian input noise and the non-linearity follows a Log-normal distribution that is closely fitted by a Gamma distribution}. (C)~This drive is finally transformed into spike counts using a Poisson point process with \xadded{the} mean \xadded{as} the input drive. The resulting tuning curve (mean spike count) is plotted in black along with the range of {$\pm$} the standard deviation (red line). The area corresponding to the mean {$\pm$} the standard deviation expected given the Poisson model {($\sigma=\sqrt \mu$)} is plotted in blue. The resulting spike counts variance is higher than what is expected by a Poisson model, as can be observed in neurophysiological data (see Figure~\ref{fig:obs} - Right).
}
\label{fig:model1}
\end{figure}
%------------------------------------------------- %
%: Gamma as lognormal
\paragraph{Over-dispersion may result from extrinsic sensory noise: \xadded  {NBM as a mixture of Poisson and LogNormal distributions.}} Many studies \xremoved{indistinctly}  assume a Log-Normal or a Gamma distribution when analyzing spike count distributions as it most often produces similar results~\citep{ Mccullagh89,Atkinson82}. Indeed, a Log-Normal process might be a good approximative model for a Gamma process. The underlying spiking process could result from a simple encoding model: Let us assume that a noisy sensory input is transformed to a response by a linear correlation measure with a cell's receptive field (for instance, in the early visual system, selective to a given orientation) and then through a non-linearity into a stochastic spike count (Figure~\ref{fig:model1}). If we consider \xremoved{an} additive extrinsic Gaussian noise as the sensory noise and an exponential function as the non-linearity, the resulting input drive follows a Log-normal distribution that is closely fitted by a Gamma distribution. Using such a model, \xremoved{the}Figure~\ref{fig:model1} shows that \xchanged{overdispersion}{over-dispersion} could simply result from an additive normal sensory noise and the non-linearity of the spiking response. 

However, even if log-Normal and Gamma distributions are often used to model the same phenomena, looking at their log functions - in general - shows clear differences in their respective skewness. Furthermore, there is no exact morphing between the two distributions that allows to link their respective parameters analytically. As such, this approximation does not allow a precise quantification of the relationship between the noise source and the \xchanged{overdispersion}{over-dispersion} parameter.

%: Gamma as sum of exponential distribution

\paragraph{Over-dispersion may result from redundancy within a neural population: \xadded  {NBM as a mixture of Poisson and Exponential distributions.}} 

The Gamma distribution could also result from a mixture of distributions. Indeed, it is well known that the sum of $n$ independent and identically distributed exponential random variables $\mathcal{E}(\frac 1\lambda)$ (the parameter of an exponential distribution is by definition the inverse of its mean, here called $\lambda$) follows a Gamma distribution $\mathcal G(n,\lambda)$, where $n$ and $\lambda$ are its shape and scale parameters respectively. As such, exponential distributions have previously been used as a simple model of variability in formal neural networks~\citep{Treves91, Levy96, Baddeley96}. They correspond to a model of the distribution of time between two Poisson events. In addition, such a distribution maximizes the entropy of spike counts (knowing the average count) and thus maximizes the carried information~\citep{Shannon48}.

%------------------------------------------------- %
%: fig:model22
\begin{figure}
\includegraphics[width=\textwidth]{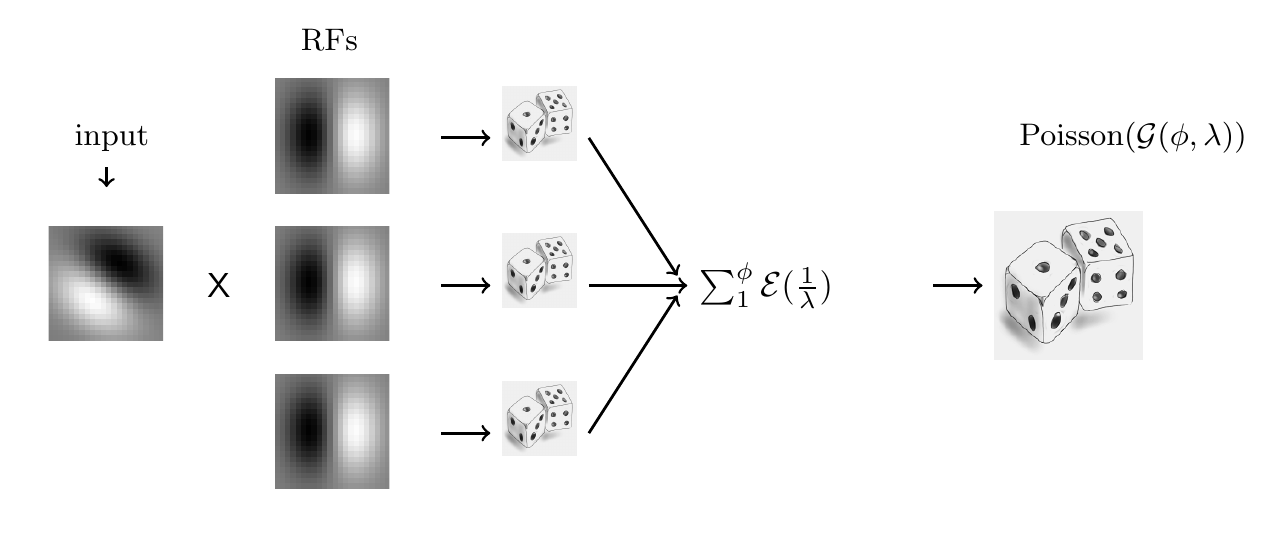}
\caption{{\bf Over-dispersion may result from redundancy within a neural population.\xadded{NBM as mixture of Poisson and Exponential distributions.}} Let us consider, within a neural population, a subset of $\phi$ cells with a similar selectivity and thus driven by a similar correlation measure $\lambda$= $\mathcal C(\theta)$. This measures the correlation between the input Gabor and the Gabor representing the orientation $\theta$ characteristic of their receptive field (RF). Let us further assume that the filter cells exhibit an exponential distribution of spike counts $\mathcal E(\frac{1}{\lambda})$ with mean $\lambda$, thus maximizing the entropy of the input drive. Then, the pooling of the resulting spike counts  results in an input drive $\mathcal G{(\phi,\lambda)}$ equal to the sum of exponential variables having the same mean equal to $\lambda$. It can be shown that the Poisson spiking process driven by this input drive is equivalently described by a Negative-Binomial probability distribution function with parameters $\lambda$ (mean) and $\phi$ (inverse-dispersion) (see equation~\ref{eq:NB}). Such a distribution describes \xchanged{well}{observed} trial-to-trial variability \xadded{well} and in this model, the parameter of dispersion is directly correlated with the redundancy, i.e., the convergence of similar input cells. }
%%%%
%The Gamma input to the Poisson generator with an inverse-dispersion parameter $\phi$ is driven by the sum of $\phi$ exponential variables having the same mean $\lambda$. These exponential variables could result from the correlation measure $\lambda(\theta)$ between the stimulus and different cells having the same filter under a multiplicative noise $-log(\mathcal U_{(0,1)})$, where $\mathcal U_{(0,1)}$ is a uniform distribution between 0 and 1. Thus, in this case, the parameter of dispersion is directly correlated with the redundancy (i.e. the convergence of similar input cells). }

\label{fig:model22}
\end{figure}
%------------------------------------------------- %

%
 From a biological point of view, to drive a Gamma input - with an inverse-dispersion parameter $\phi$ equal to $n$ - to the output neuron considered as a Poisson generator, this neuron should receive as input the sum of $n$ exponential variables having the same mean $\lambda$ (see Figure~\ref{fig:model22}). Thus,  $\phi$ reflects the amount of convergence to a neuron which is - for instance - consistent with a model of the convergent connectivity from V1 to area V5/MT~\citep{Simoncelli98, Rust06, Wang12}. In addition, it was observed that the simulation of networks based on biologically realistic parameters~\citep{Voges10neurocomp} could lead to complex dynamics, showing in particular an excess of variability~\citep{Voges12}. Thus, we could expect an \xchanged{overdispersion}{over-dispersion} in V5/MT resulting from dimensionality reduction (i.e. projecting inputs from a high dimensional space to a lower dimensional one~\citep{Haykin99}).

To summarize, we have focused on two main possible sources of \xchanged{overdispersion}{over-dispersion} in spike counts. The first, is an extrinsic source that accounts for sensory noise, \xchanged{for instance it could correspond to the thalamo-cortical pathway. The second is a source of noise that emerges from convergent inputs to a population, as one commonly sees in the cortical hierarchy, e.g. the visual processing cascade.}{ conveyed along the thalamo-cortical pathway. The second is a source of noise that emerges from convergent inputs to a population, including feedforward, lateral and feedback.} We have \xchanged{unveiled}{shown} that the convergence of inputs between two data spaces may both result in an \xchanged{overdispersion}{over-dispersion}. In the next section, we will focus on quantitatively testing evidence for this \xchanged{overdispersion}{over-dispersion}.

%But even if the Negative-Binomial model has been considered as the model of choice allowing to account for over-dispersion~\citep{Osgood00}, several studies were specifically interested on the special circumstances under which this claim is true~\citep{Berk07}.

%%------------------------------------------------- %%%

\section {The Fano Gamma test: How to quantify \xchanged{overdispersion}{over-dispersion} evidence?}
% ------------------------------------------------- %% ------------------------------------------------- %% ------------------------------------------------- %
%introduction%
 
\paragraph {The Fano factor test is misleading}  Most studies concerned with the characterization of spike count variability using different methods and formal tests start with the Fano factor (FF). However, most of the methodological studies measuring trial-to-trial variability in firing rates~\citep{Teich97, Koch99, Dayan01} do not take the estimation uncertainty introduced by the limited number of trials into account. Each trial may be considered as a realization of the random process and when accumulating evidence from $N$ trials, the average and the variance converge with an expected error of $1/\sqrt{N}$. Instead, the computed mean and variance are often treated as real features of the distribution whereas they actually correspond to estimations given the (necessarily) limited set of measurements. Thus, considering only the FF value may induce errors in conclusions regarding the assumption of probability distribution. In particular, when the observed FF deviates from $1$, is this  the result of the limited number of trials or due to an \xchanged{overdispersion}{over-dispersion}? Or when a value is measured close to $1$, can it be safely attributed to a Poisson model?

%: -previous methods
\paragraph {The chi-square test is problematic} To evaluate how likely the observed data was generated by a Poisson model, the most common approach is to use a chi-square goodness of fit test ($\chi^2$). This statistical tool can be used to estimate how closely an expected distribution matches an observed distribution of a discrete quantitative variable (having only finite possible values). However, it has been shown that small expected frequencies may result in a loss of power, that is, a tendency to not reject a false null hypothesis (or to reject a true null hypothesis)~\citep{Cochran54, Yates99, Campbell07}. This is a potential problem when we deal with small sample sizes and its outcome also depends on how data is pooled into categories. Indeed, it often makes sense to pool the small frequencies together. Moreover, \xadded{as} this test is one-tailed \xremoved{test}, \xremoved{so} it does not distinguish evidence for \xchanged{overdispersion}{over-dispersion} from evidence for \xchanged{underdispersion}{under-dispersion}.

%------------------------------------------------- %

%: Fano Gamma test 
\paragraph {The Fano Gamma test} To better quantify the \xchanged{overdispersion}{over-dispersion} while accounting for sample size and without dealing with pooling problems, we propose a statistical test based on an empirical sampling distribution of the Fano factor under the Negative-Binomial assumption. Here, we opt for an analytic procedure that we call Fano Gamma test ($FG$). It allows for a simple computation (or tabulation) of probability bounds for specific distributions hypotheses. That is not the case for other existing procedures~\citep{Amarasingham06, Jackson07, Nawrot08}. This procedure characterizes the probability distribution of the estimated FF. Thus, it allows one to compute FF bounds with respect to p-value precision limits. Starting from previous analyses~\citep{Hoel43, Kathirgamatamby53, Eden10} about the distribution of the Fisher dispersion index, we propose here that, for a Negative-Binomial distribution, the FF asymptotically follows a Gamma distribution with a shape parameter $\frac{n-1}{2}$ and a scale parameter $\frac{2(\frac\mu\phi+1)}{(n-1)}$:
\begin{equation}
 FF_{Negative-Binomial} \sim \Gamma(\frac{n-1}{2},\frac{2(\frac\mu\phi+1)}{(n-1)})
 \label{eq:FF_NB}
\end{equation}where $n$ is the number of spike count observations. We tested the accuracy of this approximation through simulation by comparing it to the empirical estimates of the true distribution of the Fano factor for any ($\mu$, $\phi$, $n$) triplet. We found that the p-value upper bounds resulting from the approximative formula are slightly over-conservative. Thus, it is worthwhile mentioning that using it to reject the NBM hypothesis is based on a strong assumption (i.e. a rejected realization could be a good one because it's p-value is between the approximation and the empirical upper bound) but we do not detail this analysis here. For large values of $\phi$, the Negative-Binomial distribution approaches a Poisson distribution and the previous formula gives: 
\begin{equation}
 FF_{Poisson} \sim \Gamma(\frac{n-1}{2},\frac{2}{n-1})
 \label{eq:FF}
\end{equation}
\citet{Eden10} studied the FF distribution under the Poisson distribution hypothesis, and have shown that the convergence to this asymptotic distribution is fast. Thus, this analysis provides a formal test  for both hypotheses (PM and NBM) given the observed variability in the spiking. 

%------------------------------------------------- %
%: fig:power
\begin{figure}
\begin{center}
\includegraphics[width=0.7\textwidth]{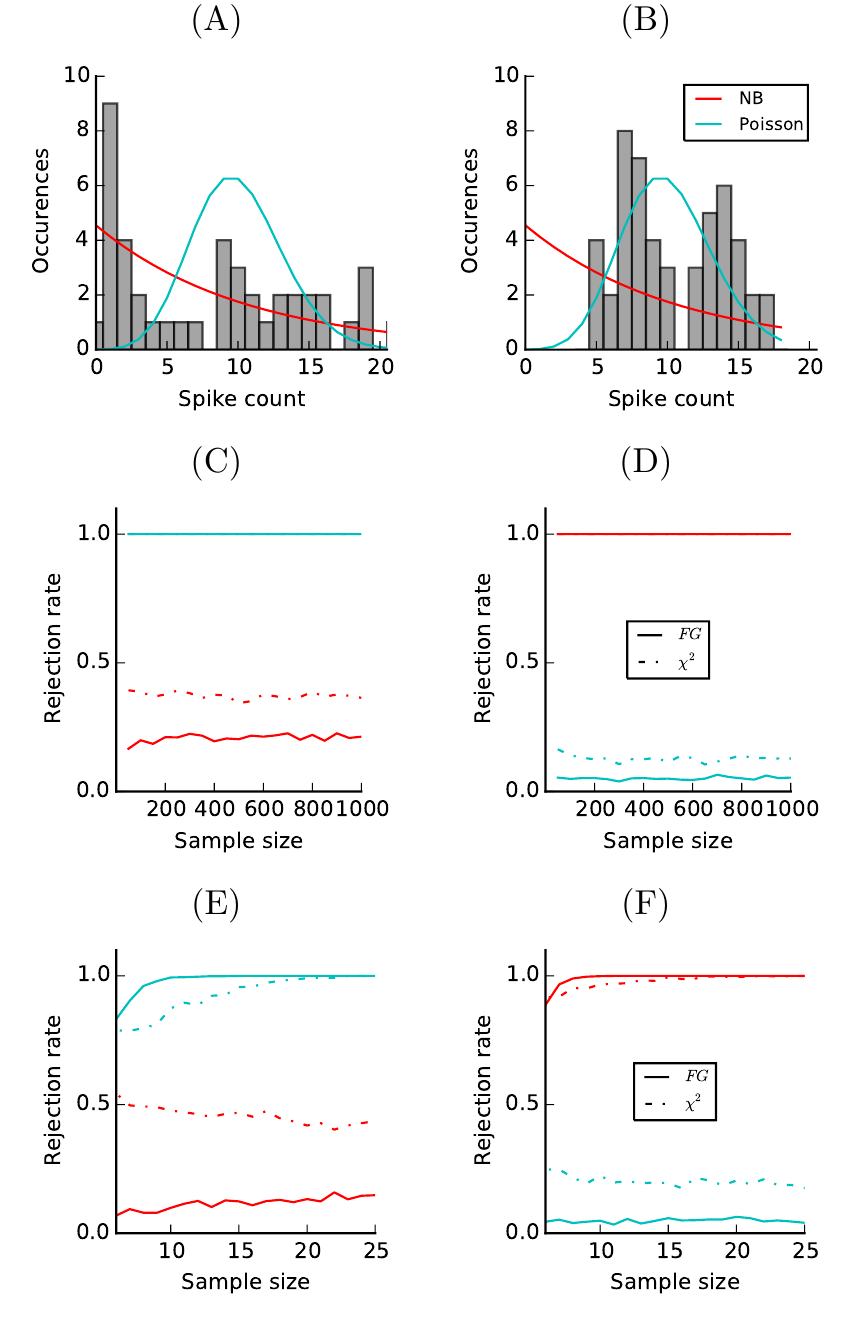}
\end{center}

\caption{{\bf Evaluating the power of chi-square test ($\chi^2$) test versus Fano Gamma test ($FG$) when testing evidence against Poisson (PM) model (in blue) and Negative-Binomial (NBM) model (in red)}. (Top) Two examples of spike count histograms resulting from artificially generated data (surrogate data) using the NBM (A) and the PM (B). These examples were chosen such that, when using the $\chi^2$ test, the model used to synthesize the data is rejected. For each histogram, we plotted in blue the expected distribution from a PM and in red the expected distribution from a NBM. For these examples, the sample size is 50. (Bottom)  $\chi^2$ test (dashed lines) and $FG$ test (solid lines) were applied to Negative-Binomial (C) and Poisson (D) surrogate data for different sample sizes with a significance level of 0.025. The mean is fixed to 10 spikes, the dispersion parameter is equal to 1 spike and the test was repeated 1000 times for each sample size. (E) and (F) show the results in (C) and (D), respectively, corresponding to small sample sizes. The rejection rate is higher for $\chi^2$  test when the null hypothesis is the correct distribution (the model generating the data) and lower for the incorrect distribution than for the proposed Fano Gamma test. The results thus show this test is more powerful than the classically used $\chi^2$ test.  }
\label{fig:power}
\end{figure}
%------------------------------------------------- %
We first compared the power of this test to the chi-square test for the general NBM hypothesis and for the specific PM hypothesis on surrogate data of spike counts. Figure~\ref{fig:power} shows that the Fano Gamma test did better than a chi-square goodness of fit test at a significance level of 0.025. This latter test has a larger probability to make a type $II$ error for both distributions (dashed red lines in Figure~\ref{fig:power}.C and dashed blue lines in Figure~\ref{fig:power}.D). 

\paragraph {Over-dispersion evidence from neurophysiological data} We applied these tests to evaluate evidence against\xadded{\footnote{Indeed, an hypothesis test is a statistical analogy to proof by contradiction. Thus, we used the  ``evidence against'' term that is the result of the statistical test and that means the rejection plausibility. Importantly, the evidence against a claim $\mathcal{H}0$ does not imply the evidence (plausibility) of the opposite claim $\mathcal{H}1$, i.e., this is not the same as a likelihood based test that allow explicitly to quantify the weight of evidence for one hypothesis.}} the Poisson model on three data sets.

The first data set results from recording \xadded{signals from} \xremoved{in} mouse LGN \xchanged{cells}{neurons} (72 cells) in response to moving \xremoved{test}gratings. The stimulus was presented several times in 12 different directions. The second data set comes from extra-cellular recordings in area V1 (67 cells) of two awake macaque monkeys in response to an oriented moving bar. The task was repeated several times for 12 different directions (orthogonal to the bar's orientation). The third data set results from recording in area MT (40 cells) of three anaesthetized macaque monkeys in response to moving gratings. The stimulus was presented several times in 16 different directions. See the methods section for more details on the data collection.

\begin{figure}
\includegraphics[width=\textwidth]{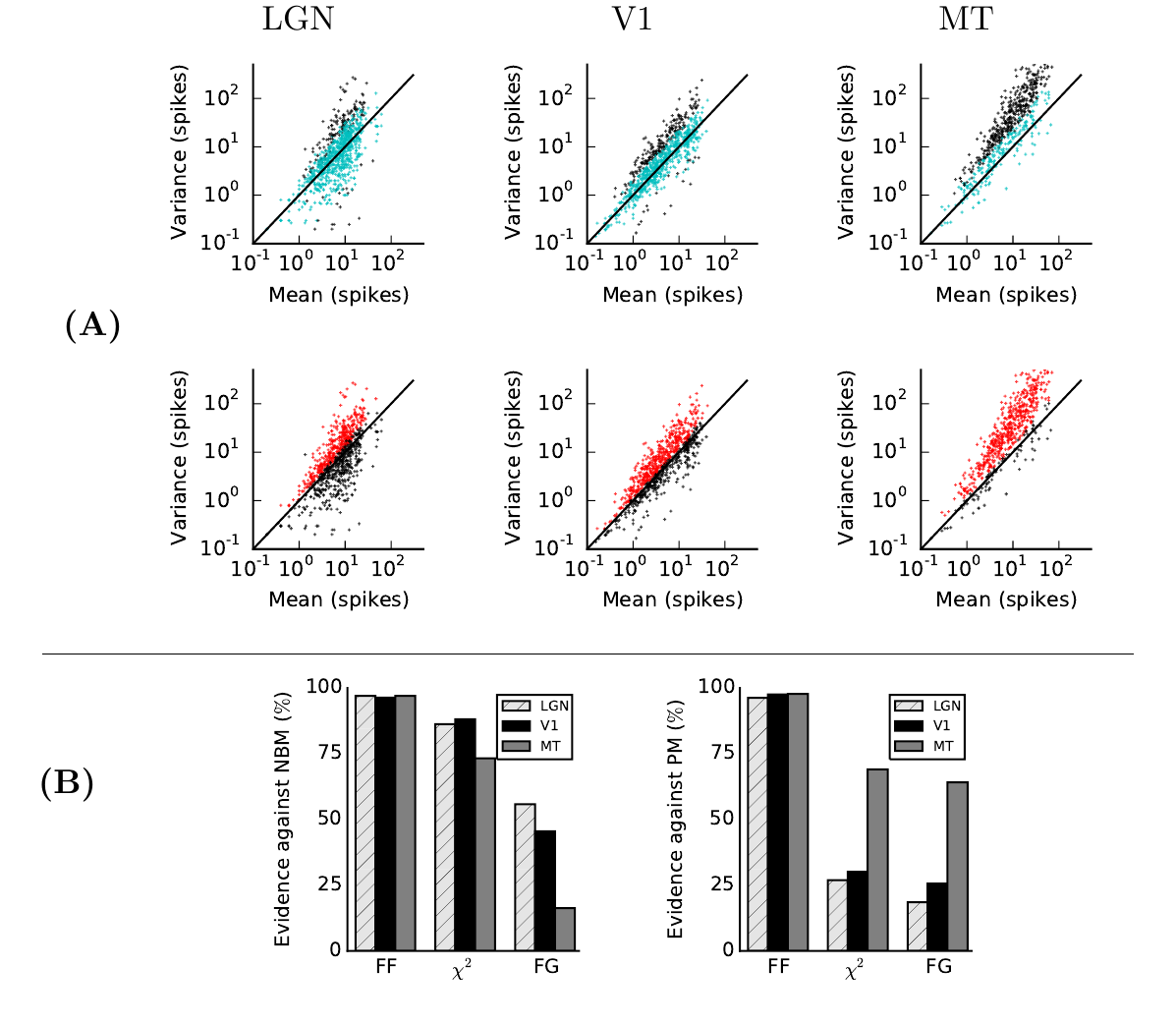}
\caption{{\bf Evaluating evidence against Poisson and Negative-Binomial models on visual neurophysiological data} (A) The Fano Gamma test (FG) results : Evidence against the Poisson model. Each subplot shows the measures of variability for the different cells and conditions as points on a (mean, variance) plot. The straight line corresponds to FF=1 (Poisson), while dots correspond to observations. Blue dots correspond to observations with no evidence against Poisson, red dots correspond to no evidence against the Negative-Binomial model (NBM) and black dots correspond to strong evidence against the Poisson model (PM) on the top and against NBM on the bottom based on the empiric distribution of the Fano factor (FG test). The results show that there is more evidence of \xchanged{overdispersion}{over-dispersion} in MT and that the proportion of cells showing \xchanged{overdispersion}{over-dispersion} seems to increase along the visual hierarchy. (B) The percentage of observations with evidence against PM and NBM are given with respect to three criteria: the Fano factor test discards the observations with a Fano factor value higher or lower than its formal expected value (equal to one for the Poisson model), the $\chi^2$ test and the $FG$ test with a significance level of 0.025. The bar plots correspond to the percentage of observations with evidence against the NBM (left) and the PM (right). The tests were applied to 3 data sets (from the left to the right, respectively): LGN cells (mouse), V1 cells (awake macaque monkey) and MT cells (anaesthetized macaque monkeys). The results show that the Fano factor values are misleading when the number of trials is not taken into account. Looking only at the FF gives strong evidence against PM and NBM that is not confirmed by the other tests. The $\chi^2$ test shows slightly more evidence against the Poisson model than the FG test which could be explained by categorization issues. It shows also very strong evidence against the NBM model known to be suitable for over dispersed data. In summary, the FG test gives more coherent results: weak evidence of \xchanged{overdispersion}{over-dispersion} in LGN and V1 data sets and low evidence against the NBM in MT data set. }
\label{fig:test}
\end{figure}

We show the results in Figure~\ref{fig:test}. For the three data sets, we found that the Fano Gamma test gives lower evidence against the PM and the NBM compared to the FF and $\chi^2$ tests (as expected given our discussion above). The Fano Gamma test shows, as do the other tests, that the evidence of \xchanged{overdispersion}{over-dispersion} is only significant in MT (with a rate of $64\%$) among cell$/$condition pairs, compared to LGN and V1 (with rates of  $19\%$ and $26\%$, respectively). Similarly, it shows that evidence against the NBM is significantly lower in the MT dataset ($15\%$) compared to LGN and V1 data sets ($52\%$ and $41\%$, respectively).
%------------------------------------------------- %
%: fig:test
%------------------------------------------------- %
%%%%%%%%%%%%%%%%%%%%%%%%%%
For these particular datasets where the number of trials is limited (mostly less than 20 trials per condition), we have shown that in the LGN and V1 data sets, the high values of Fano factor observed could be simply related to the sampling of the stochastic process itself. Indeed, looking only at the FF values induces a false belief that these datasets exhibit a significant degree of \xchanged{overdispersion}{over-dispersion}, while the FG test reliably rejects this hypothesis. The $\chi^2$ test shows slightly more evidence against the Poisson model than the FG test which could be explained by categorization issues. It shows also very strong evidence against the NBM model known to be suitable for over dispersed data. Finally, we have shown that the situation is different for the MT data set where we could indeed show evidence of a higher variability than expected from a Poisson model and less evidence against a Negative Binomial model. It is worth mentioning that our statistical test (as validated on surrogate data) allows us to make this conclusion, while other tests fail to do so. Since we mostly observe \xchanged{overdispersion}{over-dispersion} in the MT data set, we propose to compare the performance of an NBM-based to a PM-based direction decoder applied to this data set. %It also makes sense physiologically, given the Simoncelli and Heeger model~\citep{Simoncelli98}.
%%%%%%%%%%%%%%%%%%%%%%%%%%

%-------------------------------------------------------%%------------------------------- %
\section{NBM-based direction decoder: what if the \xchanged{overdispersion}{over-dispersion} is tuned ?}
To evaluate the gain in decoding when accounting for the \xchanged{overdispersion}{over-dispersion} using NBM compared to PM, we extended the classical probabilistic decoding approach proposed by~\citet{Jazayeri06} and tested the population decoder on the MT data set (presenting the highest \xchanged{overdispersion}{over-dispersion} probability). Based on a Poisson model of trial-to-trial variability applied to single neurons and a model of tuning for a population coded variable (direction $\theta$). This decoding approach allows us to infer the stimulus (the hidden variable $\theta$) given a particular neuronal response (an observed spike raster Y).

The decoding algorithm consists of maximizing the posterior probability $P({\theta}|Y)$ function of the estimated direction ${\theta}$ given a distribution hypothesis. Bayes' rule gives:  $P({\theta}|Y)=\frac{ P(Y|{\theta}) P(\theta) }{P(Y)}$, where the evidence term $P(Y)$ is a normalization term independent of ${\theta}$. So, if we assume that there is no prior knowledge on ${\theta}$ ($P(\theta)$ constant), maximizing the posterior under the PM hypothesis is equivalent to maximizing the following log-likelihood function $LP(\theta)$:
\begin{equation}
 LP(\theta) =  \sum_{i=1}^N{k_i\log[f_{i}(\theta)]}-\sum_{i=1}^N{f_{i}(\theta)}
- \sum_{i=1}^N \log[{k_i!}],\label{eq:lp}
\end{equation} where $f_{i}$ \xremoved{corresponds to the mean response of each cell $i$}is a function of the stimulus parameter $\theta$ \xadded{and corresponds to the mean response of each cell $i$}, $k_i$ is the number of trials and $N$ is the number of cells. Similarly, under the NBM hypothesis, we maximize the following log-likelihood function $LNB(\theta)$ in order to estimate ${\theta}$, where $p_i(\theta)=\frac{\phi_i(\theta)}{\phi_i(\theta)+f_i(\theta)}$ and $\phi_i$ corresponds to the inverse-dispersion parameter of each cell $i$ as a function of the stimulus parameter $\theta$: %(Equation~\ref{eq:lnb}). 
\begin{equation}
 LNB(\theta) =  \sum_{i=1}^N{\log[\frac{\Gamma(k_i+\phi_i(\theta))}{\Gamma(\phi_i(\theta))k_i!}]} +\sum_{i=1}^N {\phi_i(\theta)\log[p_i(\theta)]}+\sum_{i=1}^N {k_i\log[1-p_i(\theta)]}, \label{eq:lnb}
\end{equation} 
%\subsection*{Tuning curves}

\begin{figure}
\includegraphics[width=\textwidth]{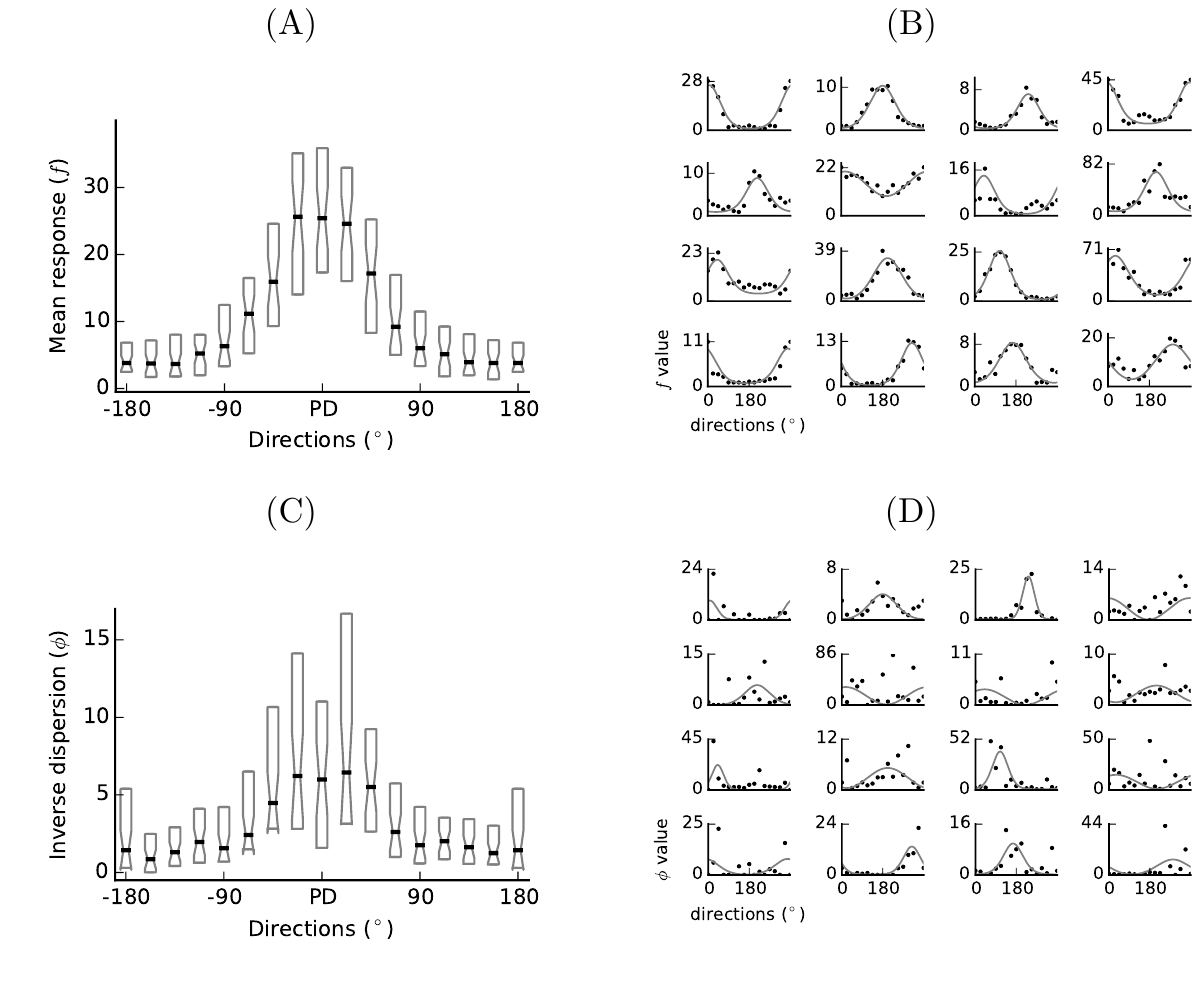}
\caption{{\bf Mean and dispersion tunings} (Left) The box plots represent the distribution of cells' mean spike count (A) and the distribution of their inverse-dispersion values (C) function of the different directions centered around the preferred direction of each cell. The ``whiskers'' for each direction indicate the 25th and the 75th percentiles. The colored dashes (blue for mean and red for inverse dispersion) represent the median values. The inverse-dispersion parameter (C) shows a tuning less regular than the mean (A) but that could, as a first approximation, be fitted by a bell-shaped function. (Right) Tuning curves of the mean spike count (B, in blue) and the inverse-dispersion (D, in red) for 16 cells  resulting from, respectively, the fit of a von Mises function (equivalent to a circular Gaussian) to each cell estimated parameters from their repeated responses to the 16 directions used for training (gray points). Each subplot correspond\xadded{s} to a cell. The tuning functions of the inverse-dispersion are supposed to be centered on the cell's preferred direction deduced from the mean fit.}
\label{fig:tunings}
\end{figure}

Given that objective, let us first define the tuning functions of the cells. These correspond to the mean spike count (the mean parameter of the Negative-Binomial model) with respect to direction. Thus, we defined for the MT data set a generative tuning function as von Mises functions peaking on the preferred direction. The parameters corresponding to the scaling, maximum and minimum values ($\kappa_o$, $\kappa_d$, $R_0$, $R_0$+$R_m$) are free to vary. The main tuning function is given by:
%s for V1 and MT are respectively given by $f_{V1}$ and 

%\begin{equation}
%f_{V1}(\theta) = R_0 + R_m \cdot \frac{e^{\kappa_o\cdot(\cos(2(\theta-\theta_0))-1)}}{e^{\kappa_o}-e^{-\kappa_o}} \cdot e^{\kappa_d\cdot(\cos(\theta-\theta_0)-1)} 
%\end{equation}

\begin{equation}
f_{MT}(\theta) = R_0 + R_m \cdot \frac{e^{\kappa_d\cdot \cos(\theta-\theta_0)}-e^{-\kappa_d}}{e^{\kappa_d}-e^{-\kappa_d}}
\end{equation}
Figure~\ref{fig:tunings}.B shows the estimated tunings of $f$ for 16 different cells.
%%---------------------------------------------------%
%: fig:tuning
%%---------------------------------------------------%

Then, we should consider the estimation of inverse-dispersion parameter $\phi$. We tested two hypotheses. First, having no prior on the tuning of $\phi$ as a function of the stimuli direction, we supposed that this parameter is stimulus independent (constant chosen as the average value of its estimates over the different conditions) and varying between cells. Second, we tested the hypothesis of a stimulus dependent inverse-dispersion parameter. We considered a von Mises tuning of $\phi$ centered on the preferred direction for each cell. Figure~\ref{fig:tunings}.D shows the estimated tunings of $\phi$ for 16 different cells. \xadded{Please note that our aim is not to have an optimal fit but to show that a simple tuning hypothesis of $\phi$ can improve the quality of decoding, even if the estimation of $\phi$ values is difficult (see discussion)}.

Finally, to perform population decoding, we used a leave-one-out (LOO) cross validation method. In this method, we randomly pick a single trial from the data set for each neuron and each condition. Then, we apply the decoding approach previously described: we infer an estimate of the encoded direction (stimulus direction) that maximizes the log-likelihood functions based on the Poisson model (eq.~\ref{eq:lp}) or the Negative-Binomial model (eq.~\ref{eq:lnb}) using the spike count of each cell given by the picked trial (as $k_i$) and the estimated values of mean and inverse-dispersion using the remaining trials(as $\phi_i$ and $f_i$). Then, we compare the estimated direction angle to the stimulus direction for the PM model and the NBM model under tuned and un-tuned dispersion hypotheses.

%%---------------------------------------------------%
%: fig:decoding
\begin{figure}
\begin{center}
\includegraphics[width=\textwidth]{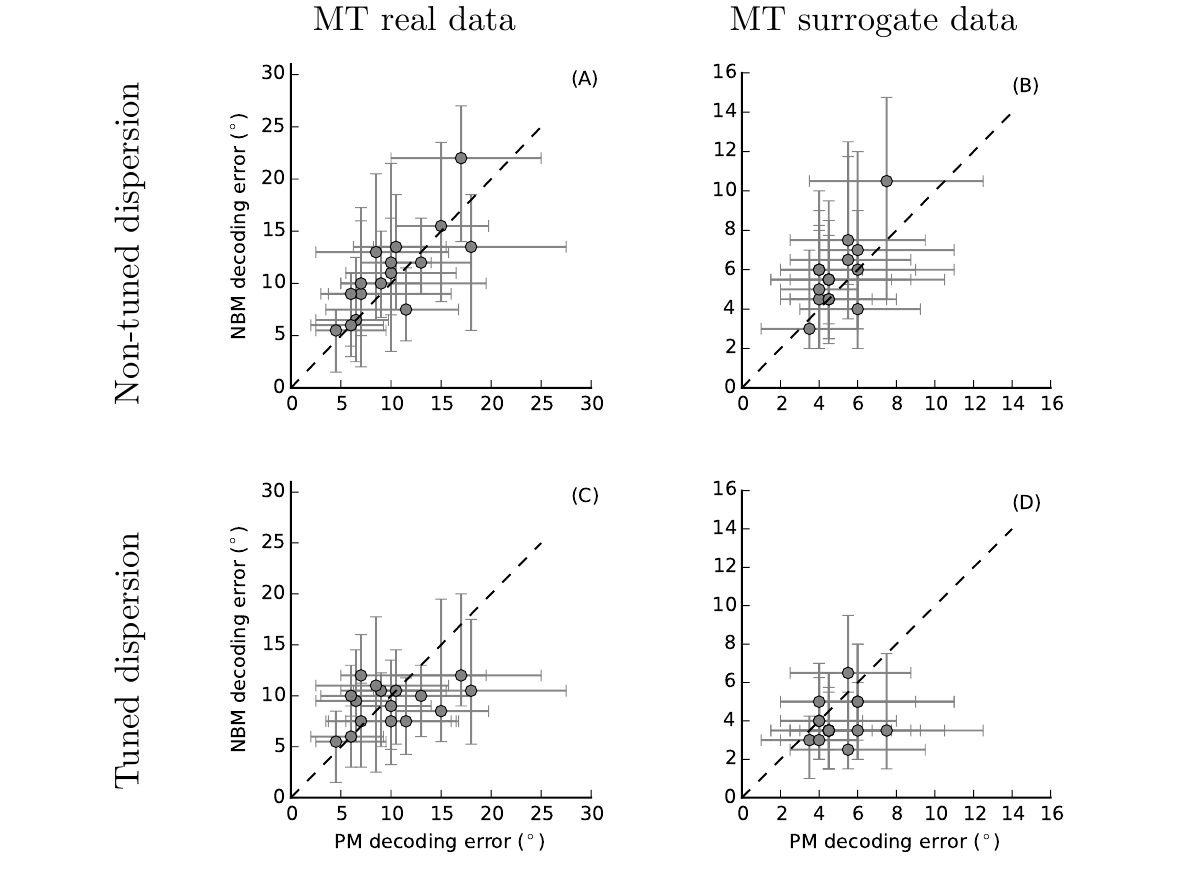}
\end{center}
\caption{{\bf Population decoding : Poisson model versus Negative-Binomial model}. Variation of decoding error values using the Negative-Binomial model (NBM) function compared to that using the Poisson model (PM) both applied to real MT data (A and C), then to surrogate data generated using the parameters inferred from the real data (B and D).  We tested two alternatives of the NBM : one with an \xchanged{overdispersion}{over-dispersion} parameter stimulus-independent (non-tuned dispersion) and the other fitting this parameter by a von Mises function of the direction centered on the preferred direction (tuned dispersion). Points correspond to the median error values and bars to 25\%-75\% confidence interval within 100 trials (horizontal for PM and vertical for NBM). The Poisson model and The Negative-\xadded{Binomial }model with non-tuned dispersion show similar results. However, the Negative-Binomial model with stimulus-dependent dispersion parameters shows better results compared to the two others. This \xremoved{shows}{suggests} that tuned \xchanged{overdispersion}{over-dispersion} could play a role in the neural code.}%
\label{fig:decoding}
\end{figure}
%%---------------------------------------------------%

The resulting decoding error values for both models applied to MT data show that the precision of decoding of the PM and the NBM under the non-tuned dispersion hypothesis are very similar (Figure~\ref{fig:decoding}.A). However, we found that there is a significant gain in decoding when using the NBM under the tuned dispersion hypothesis (Figure~\ref{fig:decoding}.C). To test if this is related to some data or estimation bias, we performed the same test on surrogate data. We generated Negative-Binomial data using the estimated parameter and applied the two decoding schemes. Similarly, the results of decoding on surrogate data show that the precision of decoding is not improved when using non-tuned dispersion (Figure~\ref{fig:decoding}.B). However, when considering additive knowledge about \xchanged{overdispersion}{over-dispersion} tuning, there is a clear gain in accuracy and precision (mean and variance Figure~\ref{fig:decoding}.D) which seems to follow decoding performance in real data.

\xadded{Moreover, comparing the goodness of fit of these models to MT data favors strongly the tuned NBM. Indeed, we used the Bayesian information criterion (BIC) that penalizes models with additional parameters \protect \citep{Box94} to determine which model best explained our data. With the BIC measure, 37 (respectively 28) over the 40 cells were best fit with the tuned NBM (respectively non-tuned NBM) and the remaining were best fit with the Poisson model.}

In summary, we have shown two main results in this section. First, the Poisson model performs as well as the Negative-Binomial model in the decoding task most of the time but not for good reasons as the data set is not well described by a Poisson variability model. Second, the Negative-Binomial model with tuned \xchanged{overdispersion}{over-dispersion} performs better and could be a generic model of neural computation.

% ------------------------------------------------- %
% ------------------------------------------------- %
% ------------------------------------------------- 
%
\section{Discussion}

%: summary%
% ------------------------------------------------- %% ------------------------------------------------- %% %------------------------------------------------- %
In this paper, we explored \xchanged{overdispersion}{over-dispersion} in neural spike counts using a doubly stochastic model, namely the Negative-Binomial model (NBM), that could disentangle different sources of \xchanged{overdispersion}{over-dispersion}. We have first explained that some commonly used procedures to quantify \xchanged{overdispersion}{over-dispersion} (the Fano factor and chi-square goodness of fit tests) could be problematic. Then, we have shown that the alternative analytical procedure that we propose (the Fano Gamma test) better characterizes \xchanged{overdispersion}{over-dispersion} in three different data sets (different animals, tasks, conditions and areas). Using the Fano Gamma test we show that there is weak evidence of \xchanged{overdispersion}{over-dispersion} in LGN and V1 data sets and low evidence against the NBM in MT data set, \xadded{especially for high mean spiking values,} a trend expected from~\citep{Goris14}. 

In the last section, we compared the performances of the traditional Poisson model to the more recent Negative-Binomial model on a population decoding algorithm (direction decoding from spike recording in MT cells of an anesthetized macaque monkey). Under the Poisson hypothesis, a linear decoding model is optimal\xadded{,} as proposed by~\citet{Jazayeri06}. We show here that this decoder is still efficient (mean decoding error $<$ 10 degrees) even though there is strong evidence against the Poisson hypothesis. This is partly due to the tuning profile of the likelihood functions: the direction corresponding to the maximum likelihood will converge toward the stimulus' direction even if the noise model is not optimal.

%In the last section, we compared the performances of the traditional Poisson model to the more recent Negative-Binomial model  on a population decoding task (direction decoding from spike recording in MT cells of an anaesthetized macaque monkey). Under the Poisson hypothesis, a linear decoding model is optimal as proposed by~\citep{Jazayeri06}. In that paper, they proposed a feed forward model implementing the computation of a decoder. This model starts from a sensory representation, use a pooling of spike counts weighted by the mean firing rates that should be merged as being the individual likelihood values. Surprisingly, we show here that this decoder is still efficient (mean decoding error $<10$ degrees). This is partly due to the tuning profile of the likelihood function, the direction corresponding to the maximum likelihood converges toward the stimulus' direction even if the noise model is not optimal.

Intuitively, one could expect that using the compound model that accounts for the observed \xchanged{overdispersion}{over-dispersion}, one should obtain better decoding results. We first considered the hypothesis of non-tuned cells' dispersion (i.e. each cell dispersion parameter is constant). The Poisson model and The Negative-Binomial model with non-tuned dispersion show similar direction decoding results, not only on real data, but also on surrogate data. However, note that, similarly to what was reported before~\citep{Goris14}, we found that the likelihood values of the estimated directions are always higher for the Negative-Binomial based decoding compared to the Poisson based decoding (not shown), but with similar bandwidths. This leads to similar decoding accuracies, despite the fact that the NBM fits better to the data. Indeed, the spike counts corresponding to the response of the cell to its preferred direction are more likely to be generated by an NBM with higher inverse-dispersion values, i.e. Poisson-like. This could explain why PM performance could be comparable to NBM performance in decoding. 
%NB non tuned is as good as Poisson
%Intuitively, one could expect that using the compound model that accounts for the over-dispersion widely observed (i.e. the Poisson process is not adapted), one should observe better decoding results. We first considered the hypothesis of non-tuned cells' dispersion (i.e. each cell dispersion parameter is constant). Indeed, that would underlie a generic neural computation which a relative invariance with stimulus directions of the information represented by the neural populations. However, the Poisson model and The Negative-Binomial model with non-tuned dispersion show similar direction decoding results not only on real data, but also on surrogate data. Similarly to what was reported before~\citep{Goris14}, we found that the likelihood values of the estimated directions are always higher for the Negative-Binomial based decoding compared to the Poisson based decoding (not shown in this paper), but with similar bandwidths. This leads to similar decoding accuracies, despite the fact that the NBM fits better to the data. Indeed, the spike counts corresponding to the response of the cell to its preferred direction are more likely to be generated by an NBM with higher inverse-dispersion values, so more Poisson-like distributions, which could explain why PM performance could be comparable to NBM performance in decoding. 
% NB with tuned is better

In addition, the estimation of the inverse-dispersion parameter is known to be difficult. Indeed, the estimation of this parameter has been widely studied in different fields~\citep{Gourieroux84, Lawless87,  Mccullagh89,  Dai13} and several estimation methods were provided such as \xadded{the} maximum likelihood method~\citep{Piegorsch90},\xadded{ or the} method of moment and maximum extended quasi-likelihood method~\citep{Clark89}. A wide range of studies have shown that the estimation of this parameter is very challenging and can be significantly biased, especially when dealing with small data sets. It is worthwhile to point out this problem here: Since neurophysiological experiments are costly and time consuming, we often deal with small data sets. Thus, having more knowledge and control about the dispersion tuning is crucial.
%%%%

Therefore, we tested a simple possible form of dispersion tuning, that is, the tuning of the inverse-dispersion having similar preferred direction as the tuning of the spike count (using a von Mises function). We found that under the hypothesis of a stimulus-dependent inverse-dispersion, the NBM performs qualitatively better.

\begin{figure}
\begin{center}
\includegraphics[width=0.9\textwidth]{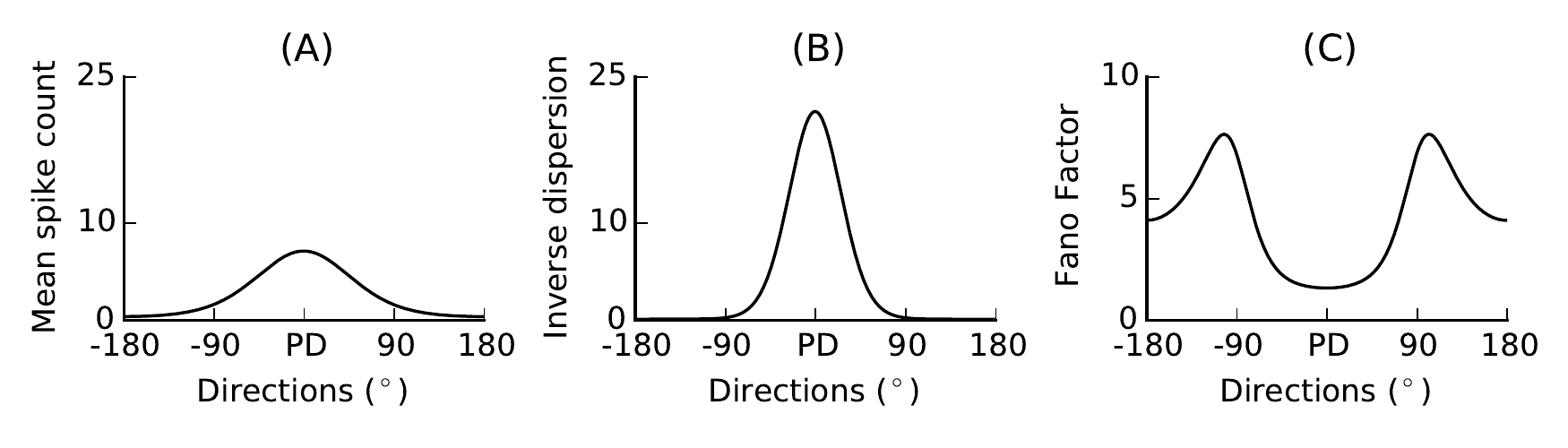}\\
\end{center}
\caption{{\bf Complex tuning of Fano factor (FF) values for different motion directions.}  \xchanged{The plot on the right shows the expected FF profile resulting from the mean spike count ($f$) profile (on the right) and the inverse-dispersion ($\phi$) profile (in the middle) using the relationship equation ($FF=1+\frac{f}{\phi}$) provided by the Negative-Binomial model.}{(A) An example of a mean response tuning curve. (B) An example of an inverse-dispersion tuning curve. (A) and (B) were analytically generated by Von Mises (circular Gaussian) functions centered on a given preferred direction (PD). (C) Let us suppose that the spiking count of a cell is driven from a Negative-Binomial model characterized by mean values given in (A) and inverse dispersion values given in (B). The expected resulting Fano factor values are given by the relationship equation ($FF=1+\frac{f}{\phi}$). For example, the mean value  (7.14) and the inverse dispersion value (21.47) corresponding to the PD  result in an FF value of 1.33. The resulting FF profile could not be explained only by the mean response tuning as in a Poisson model.}}
\label{fig:tunings2}
\end{figure}

%------------------------------------------------- %
Moreover, such a tuning function could explain complex Fano factor tuning profiles that were recently reported~\citep{Ponce13}. Our results lead us to propose that the reported trial-by-trial variability among MT neurons - showing a directional tuning that is not trivially explained by firing rate variations alone - may simply be explained by a bell-shaped tuning of the \xchanged{overdispersion}{over-dispersion} (see Figure~\ref{fig:tunings2}). \xadded{However, contrary to what they expect, we did not observe a decrease of over-dispersion with respect to the spontaneous level. Indeed, applying the Fano-Gamma test to MT spontaneous spiking data (results not shown) yields lower evidence against a PM (47.5$\%$) compared to that for evoked spiking (64.1$\%$, Figure 5.D) but similar evidence against NBM ($15\%$)  compared to evoked spiking (16.3$\%$, Figure 5.D). This is more in line with an increase of under-dispersion, but which still in accordance with the known fact that FF is decreased when a stimulus is applied.}
%%---------------------------------------------------%
%%---------------------------------------------------%

%
Where could such a tuning of the \xchanged{overdispersion}{over-dispersion} come from?  One first hypothesis could be that \xchanged{overdispersion}{over-dispersion} results from the sensory noise. Then, as suggested by~\citet{Goris14}, \xchanged{overdispersion}{over-dispersion} could also reflect an excitability state of the cell. However, both of these hypotheses cannot account for \xremoved{a}stimulus dependent variability. Alternatively, we \xremoved{can}suppose that this property is better apprehended at the population scale. Going back to the encoding paradigm described in the first section (Figure~\ref{fig:model22}), an alternative hypothesis would be that the \xchanged{overdispersion}{over-dispersion} arise from a stimulus-dependent degree of network convergence (within MT or from V1). Such \xadded{a} scheme could result from a center-surround connectivity profile in the direction domain: more excitatory input for a stimulus close to the preferred direction of the recipient cell. 
This study calls for an extension over a wider range of experimental observations. First, we would like to extend this work to more datasets with more trials to explore if our predictions are still valid, both for teasing out sources of variability but also to evaluate population decoding. It would be mostly interesting to compare this extra variability along the sensory pathways to associative areas to confirm the tendency that we observed with a gradual increase of \xchanged{overdispersion}{over-dispersion} from the lower to associative areas. A second important perspective is to consider the role of experimental conditions. Indeed, it is likely that results would greatly vary depending on the arousal \xchanged{sate}{state} of the animal, the nature of the stimulation or the behavioral task. In particular, behaviorally more relevant stimulations (such as the model-based synthesis of textures~\citep{Leon12} may lead to a more precise response~\citep{Baudot13}.%and that the tools that we developed herein could be a marker for elaborating such experiments.

%A wide range of the studies have shown that the estimation of this parameter is very challenging and can be significally biased specially when dealing with small data sets.%\citep{Willson86, Clark89, Piegorsch90, Dean94, Lord06}.

%We predict also from the FG test results on the three provided data sets that the over-dispersion is more likely to increase from low to high level areas that could be explained by  more noise sources and dimensionality reduction.
%%%%%%%%%%%%%%%%%%%%%%%%%%
\section{Materials and Methods}
In our analyses, we used three different data sets from neurophysiological recordings. 
\paragraph{LGN data set}%%%%%%%%%%%%%%%%%%%%%%%%%%
This data set~\citep{Scholl13data} contains the spiking responses of 72 LGN neurons in the anesthetized mouse to drifting grating (square wave (n=24) and sinusoidal (n=48)). Many of these neurons are already contained in the data set published in~\citep{Scholl13}. Contrarily to the two other data sets, \xremoved{this one is not from us, }it is an openly available data set. 

\paragraph{V1 data set}%%%%%%%%%%%%%%%%%%%%%%%%%%
Experiments were conducted at the INT on two adult male rhesus macaque monkeys (Macaca mulatta). Experimental protocols have been approved by the Marseille Ethical Committee in Neuroscience (approval $\#A10/01/13$, official national registration $\#71-$French Ministry of Research). All procedures complied with the French and European regulations for animal research, as well as the guidelines from the Society for Neuroscience. \\
\textit{Surgical preparation : }
Monkeys were chronically implanted with a head-holder and a recording chamber located above the V1 and V2 cortical areas. In a second surgery, a search coil was inserted below the ocular sclera to record eye movements~\citep{Robinson63}.\\
%%%%%%%%%%%%%%%%%%%%%%%%%%
\textit{Behavioral task and training : }
Monkeys were trained to fixate - within a window of 1 to 2 degrees of diameter - on a red target presented at the center of the screen during \xremoved{all}the \xadded{entire} duration of the trial (1.5 $s$). The eye position was monitored by the scleral search coil technique~\citep{Collewijn75,Robinson63} and the animal's behavior controlled using the REX package~\citep{Hays82}. \\
%%%%%%%%%%%%%%%%%%%%%%%%%%
\textit{Unit Recording :}
The two monkeys were chronically implanted with a head-holder and a recording chamber located above \xremoved{the}cortical area V1.  A computer controlled microdrive (MT-EPS, Alpha Omega, Israel) was used to trans-durally insert a micro-electrode (FHC, 0.5-1.2 M$\Omega$ at 1 KHz) in the right hemisphere. Spikes were sorted online using a template matching algorithm (MSD, Alpha Omega).\\
%%%%%%%%%%%%%%%%%%%%%%%%%%
\textit{Visual Stimulation : }
Visual stimulation protocols have been produced using in-house software (developed by G\'erard Sadoc, Acquis1-Elphy, Biologic CNRS-UNIC/ANVAR). Stimuli were back-projected on a translucent screen covering $80^\circ$$\times$$80^\circ$  of the  monkey's visual field, at a distance of 1 $m$, using a retro-projector (resolution: 1280$\times$1024 pixels at 60 $Hz$). The mean luminance of the motion stimulus was 22.2 $cd/m^2$ and the background was kept constant to about 2.24 $cd/m^2$. The display was gamma calibrated by means of a lookup table. 
Once a cell is isolated, a sparse noise (SN) stimulus was first used to quantitatively map out the RF (a grid of 10x10 squares with a side length of 0.6 degrees, 10.9 $cd/m^2$ darker or brighter than the background of 11.1 $cd/m^2$, $\sim$15 trials). Once the RF was properly located and estimated, a direction tuning paradigm was launched. In this paradigm, a moving bar (0.5$\times$4$^\circ$) translated across the receptive field (the displacement covers $3^\circ$ on either side of the RFc) in 12 possible directions (spaced by 30 degrees of polar angle, at a speed of $6.6^\circ/s$).  

\paragraph{MT data set }%%%%%%%%%%%%%%%%%%%%%%%%%%
Experiments were conducted at NYU-CNS on 3 anesthetized, paralyzed, adult macaque monkeys (Macaca nemestrina) of either sex. All procedures were conducted in compliance with the National Institute of Health Guide for the Care and Use of Laboratory Animals, and with the approval of the New York University Animal Welfare Committee.\\
%%%%%%%%%%%%%%%%%%%%%%%%%%
\textit{Surgical preparation : }The standard procedures for the surgical preparation of animals and single-unit recordings have been reported in detail previously~\citep{Cavanaugh02}. Briefly, experiments typically lasted 5 to 6 days, during which we maintained anesthesia with infusion of sufentanil citrate (6-30 $\mu g$ $kg^{-1}$ $h^{-1}$) and paralysis with infusion of vecuronium bromide (Norcuron; 0.1 $mg$ $kg^{-1}$ $h^{-1}$) in isotonic dextrose-normosol solution. We monitored vital signs (heart rate, lung pressure, end-tidal pCO2, EEG, body temperature, urine flow, and osmolarity) and maintained them within appropriate physiological ranges. Pupils were dilated with topical atropine. The eyes were protected with gas-permeable contact lenses, and refracted with supplementary lenses chosen through direct ophthalmoscopy. At the conclusion of data collection, the animal was sacrificed with an overdose of sodium pentobarbital.\\
%%%%%%%%%%%%%%%%%%%%%%%%%%
\textit{Unit Recording : }Extracellular recordings were made with quartz-platinum-tungsten microelectrodes (Thomas Recording), advanced mechanically into the brain through a craniotomy and small durotomy. To record from MT we passed microelectrodes through a small durotomy centered roughly 16 $mm$ lateral to the midline and 3 $mm$ posterior to the lip of the lunate sulcus at an angle of 20 degrees from horizontal in a ventro-anterior direction. Area MT was identified \xchanged{from}{by} the brisk direction-selective responses of isolated neurons. We made recordings from every single unit with a spike waveform that rose sufficiently above noise to be isolated. Stimuli were presented in random order.\\
%%%%%%%%%%%%%%%%%%%%%%%%%%
\textit{Visual Stimulation : }We presented visual stimuli on a gamma-corrected CRT monitor (Eizo T966; mean luminance: 33 $cd/m^2$) at a resolution of 1.280$\times$960 with a refresh rate of 120 $Hz$. Stimuli were presented using Expo software (\url{ http://corevision.cns.nyu.edu}) on an Apple Macintosh computer. For each isolated unit, we first determined its ocular dominance and occluded the non-preferred eye. We presented circularly windowed sinusoidal grating stimuli (16 directions around the clock, 22.5 degrees apart) to map each cell's receptive field, determined its preferred size and speed, and then measured selectivity for orientation or spatial frequency.

The time window considered to calculate the spike counts was the duration of stimulation for the grating stimuli (LGN and MT data sets) and 400 $ms$ covering approximatively the time when the bar stimulus \xchanged{crosses}{crossed} the cell's receptive field for V1 data set. All data were analyzed using Python and Matlab (The Mat\xadded{h}Works Inc., Natick, MA, USA). 
%%--------------------------------------------------------------------------------------------------%
\section*{Acknowledgments}
\Acknowledgments
%--------------------------------------------------------------------------------------------------%
%--%\printbibliography
%\bibliographystyle{jneurophysiol}
%\bibliography{Taouali_etal2015}
\printbibliography
%-------------------------------------------------
%\section*{Figures}
\end{document}